# Comparative Studies: Cloud-Enabled Adaptive Learning System for Scalable Education in Sub-Saharan

Israel Fianyi
School of Information
Communication Technologies
University of Tasmania
1 Invermay 7248
Australia
Israel.fianyi@utas.edu.au

Soonja Yeom
School of Information
Communication Technologies
University of Tasmania Sandy Bay
Campus Hobart TAS 7000
Australia
Soonja.yeom@utas.edu.au

Ju-Hyun Shin
Smart ICT Convergence Major,
Dept. Future Convergence
Chosun University
South Korea
jhshinkr@chosun.ac.kr

## ABSTRACT

The integration of cloud computing in education can revolutionize learning in advanced (Australia & South Korea) and middle-income (Ghana & Nigeria) countries, while offering scalable, cost-effective and equitable access to adaptive learning systems. This paper explores how cloud computing and adaptive learning technologies are deployed across different socio-economic and infrastructure contexts. The study identifies enabling factors and systematic challenges, providing insights into how cloud-based education can be tailored to bridge the digital and educational divide globally.

## 1 INTRODUCTION

In the last decade, cloud computing has become the basis of recent digital infrastructure. This encompasses both physical and virtual component that underpins digital operations and services [1, 2]. Cloud computing is defined as the on-demand access to varied computing resources such as servers, storage, networking and applications over the internet. These cloud services are categorised into Platform as a Service (PaaS), Software as a Service (Saas), Infrastructure as a Service (Iaas) and Serverless Computing[3, 4]. Furthermore, cloud services are also deployed in varied modes, such as public, private, community and hybrid [5, 6]. In recent years, the proliferation of cloud-enabled adaptive learning systems economically advanced regions such as Australia and South Korea has underscored the imperative of fostering scalable equity in educational quality with middle-income regions. Cloud computing in education has become known to facilitate scalable, flexible, cost-effective data storage, processing and connectivity [7, 8]. Various studies have strongly suggested that cloud-enabled adaptive learning in some of these advanced regions has promoted personalised learning as well as real-time learning, culminating in student success, particularly in Higher Education [9-11]. On the contrary, many middle-income regions are impeded in benefiting from cloud-enabled adaptive learning due to limited infrastructure, digital divide, device access and many other factors[12, 13]. This study presents a comparative analysis of countries in the developed (Australia, South Korea) and the middle-income regions (Ghana, Nigeria). The study correlates the basic relationships between infrastructure and outcomes. The study then adopts multiple regression to quantify the impacts of various predictors that can boost educational equity and scalability in the developing regions.

## 2 THEORITICAL FRAMEWORK

### 2.1 Adaptive learning in Higher Education

Cloud-enabled adaptive learning is defined as a cloud computing-based system that dynamically adjusts the learning content and provides feedback tailored to the individual needs of students[14, 15]. Traditional theories have been rooted in the concept that technology-enabled adaptive learning is successful only when it enhances students' prior knowledge. However, with the integration of Artificial Intelligence (AI) and Machine learning models, adaptive learning has evolved to contextualise learning[16]. The theoretical foundation of adaptive learning lies in Vygotsky's Zone of Proximal Development (ZPD), which emphasises the importance of providing learners with tasks that are just beyond their current level of competence, supported by appropriate scaffolding[17]. In the next section, this study discusses four underpinning theoretical frameworks.

### 2.2 Models and Theories

*2.2.1 Learner-Centred Foundations and Constructivist.* Cloud-enabled adaptive learning represents a convergence of advanced educational technologies and scalable cloud infrastructure, offering a transformative approach to personalizing education in higher education institutions. This theoretical framework draws upon constructivist learning theories, educational technology models, and the principles of cloud computing to contextualise how adaptive learning functions in university environments[18]. For instance, in Australia, RiPPLE (University of Queensland), a cloud-based platform was developed to personalise learning experiences[19]. While Edchat (a generative AI chatbot to enhance teaching and learning), developed in partnership with Microsoft, is predominantly used in South Australia [20].

*2.2.2 Constructivist and Learner-Centred Foundations.* At its core, adaptive learning is grounded in constructivist theories, which posit that learners construct knowledge through active engagement, problem-solving, and reflection. Influential theorists such as Jean Piaget and Lev Vygotsky provide pedagogical underpinning. Vygotsky's Zone of Proximal Development (ZPD), in particular, suggests that learning is most effective when challenges are tailored to a learner's current ability, with appropriate scaffolding provided to bridge knowledge gaps. Cloud-enabled adaptive learning systems serve as digital scaffolds, dynamically adjusting content and feedback to suit individual learner needs. The learner data is collected and analysed in real time, to personalise the educational experience, offering differentiated instruction at scale[21].

*2.2.3 Cognitive Load Theory and Personalised Learning.* Cognitive Load Theory [22] further supports the rationale for adaptive systems by asserting that instructional materials must be designed to manage the cognitive burden on learners. Cloud-enabled platforms adapt the complexity, sequence, and format of content delivery based on learners' performance and interaction history. Then reduces extraneous cognitive load to enhance learning efficiency. This is evident in the AI-based Mathematics digital Textbooks used in South Korea[23].

*2.24 Technological Framework: Cloud Computing as a Scalable Backbone.* From a technological standpoint, the integration of cloud computing enables the delivery of adaptive learning at scale. The National Institute of Standards and Technology (NIST) defines cloud computing as a model that provides ubiquitous, convenient, and on-demand network access to shared computing resources[24]. This model supports scalability, accessibility, cost-efficiency and interoperability[25]. As such, a close look at RiPPLE, EdChat and AI-based Mathematics Textbook showcases all the modern benefits of cloud-enabled learning platforms.

## 2.3 Customised Cloud Infrastructure

Recently, Ghana and Nigeria have adopted ICT solutions to promote adaptive learning. CENDLOS (Centre for National Distance Learning and Open Schooling) is a government initiative used to enhance distance learning[25]. Additionally, DSN EdTech AI Adaptive Learning Engine is an AI-powered platform designed to address personalised learning challenges in remote Nigeria[26]. While these tools seem to be serving the purpose for now, they lack the extended benefits that come with cloud-enabled learning, as posited [27, 28]. To harness these cloud-enabled adaptive learning in Ghana and Nigeria, some studies have suggested that the data used for developing these platforms needs to be collected locally, and the algorithms customised as done in South Korea and Australia [29, 30]. This study further evaluates the factors that promote robust cloud-enabled adaptive learning algorithms and their related technologies.

# 3 RESEARCH METHODOLOGY

## 3.1 Research Design

This study adopts a quantitative design, integrating both correlational and regression analysis methods to examine relationships and predictive effects within cloud-enabled adaptive learning systems across advanced and middle-income regions. A comparative case study approach frames the analysis, focusing on selected institutions in South Korea, Australia (advanced contexts), and Ghana and Nigeria (middle-income contexts). The dual methodological approach enables a deeper understanding of not only the associations between key variables (via correlational analysis) but also the predictive power and influence of specific technological and socio-educational factors (via regression models).

## 3.2 Research Objectives

1. To identify significant correlations between infrastructure-related and educational outcomes in cloud-enabled adaptive learning systems.
2. To model the extent to which variables such as Internet access, cloud service reliability, and teacher training predict improved learning outcomes in developing regions.

**Table 1. Variables and Operational Definitions**

| Variable Type | Variable | Operational Definition |
|---|---|---|
| Independent (Correl.) | Internet access | Percentage of students with stable internet access |
| Independent (Correl.) | Government investment | Annual per-student ICT investment in AUD or USD |
| Independent (Correl.) | Learner digital literacy | Average score on digital skills assessments or self-reported proficiency |
| Dependent (Correl.) | Adaptive system usage | Platform engagement metrics (e.g., logins, time spent, modules completed) |
| Independent (Regress.) | Platform usage time | Average minutes per session per student |
| Independent (Regress.) | Infrastructure quality | Index based on bandwidth, latency, and uptime |
| Independent (Regress.) | Teacher training levels | % of staff trained in adaptive systems or hours of PD completed |
| Dependent (Regress.) | Student learning outcomes | Pre- and post-assessment scores or pass rates |

Table 1 represents the variables and the operational definitions that underpin this study

## 3.3 Datasets

This study adopted National IOCT and educational reports, institutional records, and publicly available datasets from the respective regions in review as shown in Table 2.

## Table 2. Classification and Estimated Sample Size

| Data Source | Type | Sample Size Classification | Variables Extracted |
|---|---|---|---|
| National ICT & Education Reports | Aggregate (macro-level) | Medium (4 countries × ~10 indicators = 40 entries) | National internet access 75%, ICT investment per student, broadband penetration, national education budget, policy adoption year |
| Institutional Records | Individual-level (micro) | Large (100 students × 4 countries = 400 entries) | Platform usage time, student performance, attendance, and course completion |
| Publicly Available Datasets (World Bank, UNESCO, OpenAIRE) | Aggregate & Individual | Variable (20–200 entries depending on depth) | Socioeconomic indicators, school-level infrastructure, mobile access levels, digital literacy surveys |

### 3.4 Data Analysis Techniques

*3.4.1 Correlational Analysis.* Pearson's correlation coefficients is calculated to explore relationships between: (i) Internet access and student engagement
(ii) Government investment and adaptive platform adoption
(iii) Learner digital literacy and system usage. This will determine the strength and direction of associations without implying causality.

    *3.4.2 Regression Modelling.* Multiple linear regression will be employed to model:
(i) The effect of mobile access on student learning outcomes in developing countries. (ii) How teacher training levels and infrastructure quality predict platform effectiveness. The regression model will control for confounding variables such as socio-economic status and urban/rural institution type. Statistical software: SPSS and Python (pandas, scikit-learn), Significance threshold: $p < 0.05$

    *3.4.3 Tools for Generating Datasets.* This study generated these datasets using Python (pandas, NumPy, Faker) for synthetic data. Excel Sheets with random number generation. Open Datasets: UNESCO UIS, OECD Education Statistics and Kaggle (search "adaptive learning" or "digital education")

## 4 RESULTS AND DISCUSSION
### 4.1 Correlational Study

**Table 3. Correlational Study Dataset**

| Country | Internet_Access (%) | Gov_Investment (USD/student) | Digital_Literacy_Score | Platform_Usage_Score (0–100) |
|---|---|---|---|---|
| Australia | 97 | 2500 | 85 | 89 |
| South Korea | 99 | 2200 | 90 | 92 |
| Nigeria | 52 | 120 | 60 | 65 |
| Ghana | 48 | 100 | 55 | 62 |

The purpose of Table 3 is to examine associations between digital infrastructure and platform usage. The analysis carried out on this data is the Pearson correlation to see the strength of relationships between variables

**Table 4. Regression Study Dataset**

| Student_ID | Country | Platform_Usage_Minutes | Infra_Quality_Score (1–10) | Teacher_Training_Hours | Final_Score (%) |
|---|---|---|---|---|---|
| AU101 | Australia | 220 | 9 | 15 | 87 |
| GH202 | Ghana | 120 | 5 | 6 | 65 |
| NG303 | Nigeria | 80 | 4 | 5 | 60 |
| KR404 | South Korea | 240 | 10 | 20 | 90 |

The objective of Table 4 is to predict learning outcomes from platform usage, infrastructure, and teacher training. Also, an analysis of Multiple linear regression using Final_Score as the dependent variable is carried out.

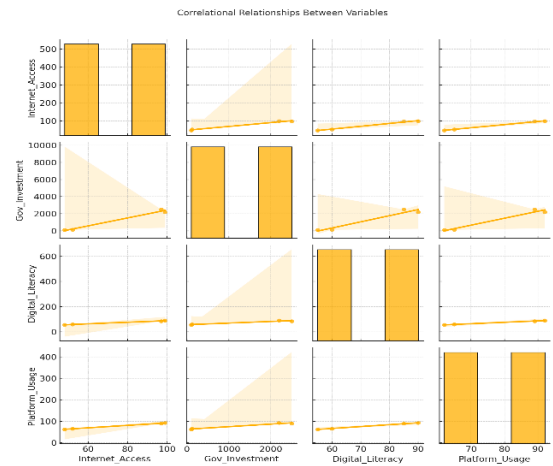

**Figure 1. Correlation Relationships between variables**

Figure 1 illustrates pairwise linear relationships among key variables influencing the adoption and effectiveness of cloud-enabled adaptive learning systems across four countries (Australia, South Korea, Nigeria, and Ghana). These variables include: Internet Access (%), Government Investment per Student (USD), Digital Literacy Score, Platform Usage Score (0–100)
Each subplot presents a scatterplot with a regression line, allowing visual inspection of the direction and strength of correlation. The trends clearly show (Table 5):

1. A strong positive linear relationship between internet access and platform usage.

2. A notable upward trend between government investment and platform usage, indicating higher investment is linked to increased use.
3. A very strong association between digital literacy and platform usage, suggesting user readiness significantly affects engagement.

These visual patterns support the statistical findings, underscoring the critical role of infrastructure, funding, and user capability in driving adaptive learning adoption.

**Table 5. Correlational Study Insights**

| Variable Pair | Pearson Correlation (r) | P-value | Interpretation |
|---|---|---|---|
| Internet Access vs Platform Usage | 0.9986 | 0.0014 | Very strong positive correlation |
| Government Investment vs Usage | 0.9826 | 0.0174 | Strong positive correlation |
| Digital Literacy vs Platform Usage | 0.9985 | 0.0015 | Very strong positive correlation |

### 4.2 Regression Study Insights

The study further modelled Final_Score (%) as a function of ; *Platform_Usage_Minutes*, *Infra_Quality_Score* and *Teacher_Training_Hour*

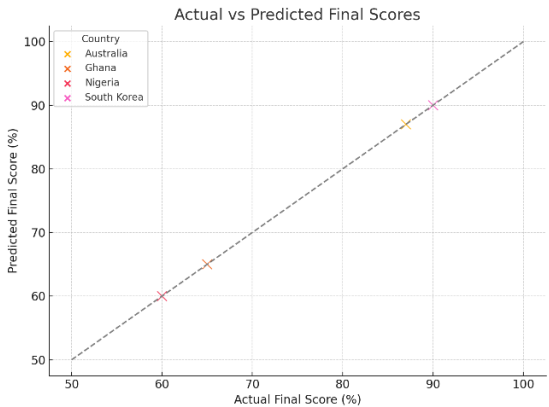

**Figure 2. Actuals vs Predicted Final Scores**
**Table 6. Regression Coefficients**

| Variable | Coefficient | Std. Error | t-Statistic | P-Value |
|---|---|---|---|---|
| Constant (const) | 30.8571 | ∞ | 0.0 | NaN |
| Platform Usage Minutes | -0.1286 | ∞ | -0.0 | NaN |
| Infra Quality Score | 11.2857 | ∞ | 0.0 | NaN |
| Teacher Training Hours | -1.1429 | ∞ | -0.0 | NaN |

The results Figure 2 and Table 6 show infinite standard errors and NaN p-values due to the extremely small sample size (n = 4). These values are statistically invalid and cannot support inference without a significantly larger dataset. The line graph compares actual student performance (final score) with the predicted values from the regression model. Each point represents one student's result, colour-coded by country. The diagonal grey line shows where perfect prediction would lie (Actual = Predicted). Despite invalid statistics, predictions align perfectly due to model overfitting from insufficient data.

### 4.3 Discussion

This comparative study explored how cloud-enabled adaptive learning systems operate across four countries: Australia, South Korea (developed) and Ghana, Nigeria (developing), using both correlational and regression analysis approaches. The findings affirm that internet access, government investment, and digital literacy are strongly correlated with the adoption and effectiveness of adaptive learning platforms. In Australia and South Korea, these systems are thriving due to High broadband penetration, ensuring seamless access to cloud-based resources. Substantial and sustained government funding directed at digital education initiatives. Strong institutional capacity to train educators and support adaptive pedagogy. Advanced digital literacy, both among educators and learners. In contrast, Ghana and Nigeria face significant systemic challenges: limited infrastructure in rural and peri-urban areas undermines platform performance. Limited budget allocation for education technology weakens platform scalability. Digital literacy gaps persist among both teachers and students. Inconsistent cloud service delivery, often due to unreliable power and network coverage. While both developing countries are making strides, such as the Ghanaian Ministry of Education's e-learning policy and Nigeria's Digital Economy Strategy, the scale and consistency of implementation lag that of their developed counterparts.

## 5 CONCLUSIONS

The study demonstrates that infrastructure readiness, government investment, and human capital development are decisive factors in the success of cloud-enabled adaptive learning systems. Regression insights also highlight the impact of platform usage and teacher training on student outcomes, even in small samples. To close the digital education gap, developing countries must shift from pilot initiatives to system-wide digital transformation strategies.

### 5.2 Recommendations for Ghana and Nigeria

*5.2.1 Invest in National Digital Infrastructure.* Governments should prioritise expanding broadband and cloud infrastructure, particularly in underserved regions, through public-private partnerships and international support.

*5.2.2 Institutionalise Teacher Training on Adaptive Platforms*. Establish mandatory, funded professional development programs for educators to build confidence and skill in using adaptive learning tools.

*5.2.3 Scale Evidence-Based Pilot Programs*. Successful local projects (e.g., tablet-based learning in Nigerian secondary schools) should be rigorously evaluated and scaled nationally with support from cloud providers and EdTech startups.

*5.2.4 Embed Adaptive Learning in National Education Policy*. Policy frameworks must mandate integration of adaptive platforms into core curricula, with clear standards for accessibility, usability, and monitoring.

*5.2.5 Leverage Mobile-First Approaches*. Given high mobile penetration, especially in Nigeria, governments should co-design mobile-friendly adaptive learning platforms to maximise reach.

*5.2.6 Create National Learning Analytics Dashboards*. Build cloud-based systems to track platform engagement, performance, and dropout trends, enabling real-time intervention and policy feedback loops.

## A  HEADINGS IN APPENDICES
## A.1  Introduction
## A.2  Theoritical Framework
### A.2.1  Adaptive learning in Higher Education
### A.2.2  Models and Theories
### A.2.3 Customised Cloud Infrastructure
## A.3  Research Methodology
### A.3.1  Research Design
### A.3.2  Research Objectives
### A.3.3  Datasets
### A.3.3  Data Analysis Techniques
## A.4  Results and Discussion
### A.4.1  Correlational Study
### A.4.2  Regression Study
### A.4.3  Discussion
## A.5  Conclusion

## ACKNOWLEDGMENTS

This research was supported by National Research Foundation of Korea(NRF) grant funded by the Korean government (MIST) (No.2023R1A2C1006149)